\documentclass[11pt,a4paper]{article}

\RequirePackage[l2tabu, orthodox]{nag}

\usepackage{jheppub}
\usepackage{amsthm,amssymb,amsmath,epic,float}
\usepackage{rotating,epsfig,indentfirst,array,varioref}
\usepackage{appendix,tikz,mathtools}
\usepackage{setspace}
\usepackage{tikz}
\usetikzlibrary{decorations.pathreplacing}
\usetikzlibrary{calc}
\usepackage{enumitem}
\usepackage{hyphenat}
\usetikzlibrary{positioning}
\usepackage{graphicx}  
\usepackage{adjustbox}

\usepackage{arydshln}

\def\be{\begin{equation}}
\def\ee{\end{equation}}

\def\bbt{\bibitem}
\def\be{\begin{equation}}
\def\en{\end{equation}}
\def\ber{\begin{eqnarray}}
\def\enr{\end{eqnarray}}
\def\nmb{ \nonumber\\}
\def\d{\partial}
\def\rbr{\rbrack}
\def\lbr{\lbrack}
\def\rbrc{\rbrace}
\def\lbrc{\lbrace}
\def\ov{\over }
\def\tld{\tilde}

\def\sgm{\sigma}

\def\bet{\beta}
\def\gm{\gamma}

\def\im{\imath}

\def\lm{\lambda}

\def\om{\omega}

\def\dlt{\delta}

\def\vl{\vec{l}}

\def\vq{\vec{q}}

\def\fr12{\frac{1}{2}}
\def\bl{\bf{l}}
\def\bt{\bf{t}}
\def\bw{\bf{w}}
\def\bk{\bf{k}}
\def\bq{\bf{q}}

\usepackage{jheppub}
\makeatletter
\def\@fpheader{\vspace{-.1cm}}
\makeatother

\title{\boldmath  Mirror symmetry and  new approach to constructing orbifolds of Gepner models\\}

\author[a,b]{Alexander Belavin,}
\author [a]{Sergey Parkhomenko}

\affiliation[a]{Landau Institute for Theoretical Physics, 142432 Chernogolovka, Russia}
\affiliation[b]{Kharkevich Institute for Information Transmission Problems, 127994 Moscow, Russia}

\emailAdd{sashabelavin@gmail.com}
\emailAdd{spark@itp.ac.ru}

\abstract{ Motivated  by the principles of the conformal bootstrap, primarily the principle of Locality, simultaneously with the requirement of space-time supersymmetry, we reconsider constructions of  compactified superstring models. Starting from    requirements of space-time supersymmetry and mutual locality, we construct a complete set of physical fields of orbifolds of Gepner models. To technically implement this, we use spectral flow generators to construct all physical fields from  the chiral primary fields. The set of these spectral flow operators forms a so-called admissible group  $G_{adm}$, which defines a given orbifold.
The action of these operators produces a collection of physical fields consistent with the action of supersymmetry generators. The selection of mutually local fields from this collection is carried out using the mirror group $G^*_{adm}$.  The permutation of $G_{adm}$ and $G^*_{adm}$ replaces the original orbifold with a mirror one that satisfies the same conditions as the original one. This also implies that the resulting model is modular invariant.}


\begin{document} 
\maketitle
\flushbottom

\section{Introduction}
\label{sec:intro}

Ten-dimensional superstring theory naturally unifies gauge quantum field theory and quantum gravity.
In \cite{CHSW} it was shown that to obtain a 4-dimensional theory with the space-time supersymmetry required for phenomenological reasons, we must compactify 6 of the 10 dimensions on the so-called Calabi-Yau manifold.
Another equivalent way to do the same is the compactification
of $6$ dimensions on N = 2 superconformal field theory with the
central charge equal $9$ \cite{Gep}.

Each of these two equivalent ways leading to the same theory has its own advantages. The first  gives a geometrical interpretation.
The second allows one to use exactly solvable N=2 SCFT models and thus obtain an explicit solution to the problem under consideration.
In this article we follow the second way and propose a new approach to constructing orbifolds of Gepner models built using the tensor product
of $N=2$ minimal models \cite{Gep}.
This approach leads to the same results as in \cite{Gep,GrPl}, based on the simultaneous fulfillment of the requirements of supersymmetry and modular invariance. But our initial requirements are different.
Namely, instead it requires simultaneous fulfillment of  space-time supersymmetry and mutual locality for construction of the space of all physical states of the theory.

The construction consists of the following  steps. As in \cite{Gep, BLT, BP, BBP, P}, we take for the  compactification the product of $N=2$ minimal models with the full central charge of $9$ joined  with  four free bosonic $X^{\mu}$, four fermionic fields $\psi^{\mu}$, as well as with BRST ghost fields.

The requirement of consistency with the action of the complete $N=2$ Virasoro algebra, which is a subalgebra of the product of $N=2$ Virasoro algebras of all components of the theory, imposes certain restrictions on the states of subtheories \cite{Gep}.
In particular, fields from the left or right $NS$ sectors of the theory must include  products of only fields from the $NS$, not $R$ sectors 
of sub-theories. The same should be done for fields from the left or 
right $R$ sectors. Then we find BRST-invariant massless fields in both the left and right sectors and select from them those that are necessary to construct space-time supersymmetry generators in both sectors. Not all physical, i.e. BRST-invariant fields, are compatible with the action of supersymmetry generators, or, equivalently, mutually local with respect to them. We choose those physical fields that are mutually local with respect to supersymmetry generators. It is equivalent to the GSO reduction.

The next step is to construct complete vertices, i.e. products of left 
(holomorphic) and right (anti-holomorphic), vertices of type $(NS,NS)$, which are mutually local. Such properties are possessed by fields $(NS,NS)$ that satisfy the GSO conditions and are diagonal with respect to the sets of their left and right $U(1)$ charges. However, since the space of all states must allow the independent action of the left and the right supersymmetry generators, it is also necessary to have twisted full vertices. That is, we need to admit such full vertices,  
whose sets of $U(1)$ charges of the  left and  right factors do not coincide.  
It is also required that the GSO conditions are satisfied in both the left and right factors, as before. To achieve the last requirement, as we will show below, it is necessary that the twisting between  $U(1)$ charge components in the left and right factors is restricted by the actions of generators forming the so-called
admissible group $G_{adm }$. Technically, we implement the action of these generators using  the spectral flow transformation \cite {SS, BP, BBP}.
At last, we check the mutual locality of the vertices constructed in this way and leave only those that satisfy this requirement.
The above-mentioned use of the spectral flow deformation turns out to be very convenient for performing this selection.
It is remarkable how it turns out that the selection of mutually local vertices is carried out using the mirror group $G^*_{adm}$.
By definition, the elements of the group $G^*_{adm}$ form the set of all generators that are mutually local with respect to the generators of the group $G_{adm}$.
After this, applying space-time SUSY generators to the resulting vertices of $(NS,NS)$ type independently in the left and right sectors, we obtain the set of vertices of $(NS,R)$, $(R,NS)$  and $(R,R)$ types. 
It is checked that the resulting vertices of all four types are mutually local. Moreover, the resulting set of physical states of the orbifolds of Gepner models coincides with that obtained earlier in in \cite{Gep,GrPl}.
Moreover, this also implies the modular invariance of the resulting model.
\section{Preliminaries}\label{sec:1}
We consider type $II$ superstring theory, whose matter sector is a product of spacetime $M_{st}$ and compact $M_{int}$ sectors.
The  space-time degrees of freedom of string in $4$-dimensional Minkowski space-time are bosons $X^{\mu}(z)$, and fermions $\psi^{\mu}(z)$, $\mu=0,...,3$ with the OPE 
\be
\begin{aligned}
&X^{\mu}(z)X^{\nu}(0)=-\eta^{\mu\nu}\log{z}+..., 
\nmb
&\psi^{\mu}(z)\psi^{\nu}(0)=\eta^{\mu\nu}z^{-1}+...,
\label{1.XpsiOPE}
\end{aligned}
\ee
and endowed with the action of $N=1$ Virasoro superalgebra with central charge $c_{st}=6$. The currents of the algebra are given by
\begin{equation*}
\begin{aligned}
&T_{st}(z)=-\frac{1}{2}\d X^{\mu}(z) \d X_{\mu}(z)-\frac{1}{2}\psi^{\mu}(z)\d\psi_{\mu}(z),\\ 
&G_{st}(z)=\im\psi^{\mu}(z)\d X_{\mu}(z).
\label{1.LVirst}
\end{aligned}
\end{equation*}

It is convenient to bosonize fermions by introducing
fields  $H_{a}(z)$, $a=1,2$ as
\begin{equation}
H_{a}(z)H_{b}(0)=-\dlt_{ab}\log{(z)}+....
\label{1.fermbos}
\end{equation}
\be
\begin{aligned}
&\frac{1}{\sqrt{2}}(\pm\psi^{0}+\psi^{1})=\exp{[\pm\im H_{1}]},
\nmb
&\frac{1}{\sqrt{2}}(\psi^{2}\pm\im\psi^{3})=\exp{[\pm\im H_{2}]}.
\label{1.ferm}
\end{aligned}
\en
Then the left part of  the vertex operators  of the space-time sector can be written as
\ber
V_{st}(z)=P_{st}(\d X^{\mu},\d H_{a})\exp{[\im\lm^{a}H_{a}]}\exp{[\im k_{\mu}X^{\mu}(z)]},
\label{1.LVert}
\enr
where $P_{st}(\d X^{\mu},\d H_{a})$ is a polynomial of $\d^{n}X^{\mu}(z)$, $\d^{m}H_{a}(z)$ and $\lm=(\lm^{1},\lm^{2})$ is the weight of $SO(1,3)$ algebra.

In $NS$ sector the weights $\lm$ fall into two conjugacy classes. The class of trivial representation $[0]$ contains the $\lm$'s which are 
the $SO(1,3)$ root lattice vectors. The class of fundamental vector representation $[v]$ contains the $\lm$'s which are the $SO(1,3)$ root lattice vectors shifted by the highest weight vector $\om_{1}=(1,0)$. 

The weights $\lm$ in $R$ sector also fall into other two conjugacy classes. The class of chiral spinor representation $[\sigma]$ contains the $\lm$'s which are the $SO(1,3)$ root lattice vectors shifted by the highest weight $\om_{2}=(\fr12,\fr12)$. The class of anti-chiral spinor representation 
$[\dot{\sigma}]$ contains the $\lm$'s which are the $SO(1,3)$ root lattice vectors shifted by the highest weight vector $\om_{3}=(\fr12,-\fr12)$.
Completely the same picture takes  place in the right-moving part of space-time sector.
 
The space-time, compact and ghost sectors of supersting theory have the 
$N=2$ superconformal symmetry.
Recall the commutation relations for the  $N=2$ Super-Virasoro algebra
\begin{equation}
\begin{aligned}
&\lbr  L_{n},L_{m}\rbr=(n-m)L_{n+m}+{c\ov 12}(n^{3}-n)\dlt_{n+m,0},\\
&\lbr J_{n},J_{m}\rbr={c\ov 3}n\dlt_{n+m,0},\\
&\lbr L_{n},J_{m}\rbr=-mJ_{n+m},\\
&\lbrc G^{+}_{r},G^{-}_{s}\rbrc=L_{r+s}+{r-s\ov 2}J_{r+s}+{c\ov 6}(r^{2}-{1\ov 4})\dlt_{r+s,0},\\
&\lbr L_{n},G^{\pm}_{r}\rbr=({n\ov 2}-r)G^{\pm}_{r+n},\\
&\lbr J_{n},G^{\pm}_{r}\rbr=\pm G^{\pm}_{r+n}.
\label{1.N2Vir}.
\end{aligned}
\end{equation}
The $L_{n}$ generators are the Fourier components of the stress-energy tensor $T(z)$, $J_{n}$ are  the modes of $U(1)$ current $J(z)$ and $G^{\pm}_{r}$ are the modes of the fermionic currents $G^{\pm}(z)$. 
Note that the  $N=1$ Virasoro subalgebra is spanned by
$L_{n}$ and $G_r=G^{+}_{r}+G^{-}_{r}$ generators.
In the compactification of type IIA/B superstrings, the  $N=2$ Virasoro superalgebra acts in the left and right sectors, so we have two copies of the algebra (\ref{1.N2Vir}). 

The one-parametric family of automorphisms of $N=2$ Virasoro superalgebra \cite{SS}
\begin{equation}
	\begin{aligned}
		&\tilde{G}^{\pm}_{r}=U^{-t}G^{\pm}_{r}U^{t}=G^{\pm}_{r\pm t},\\
		&\tilde{J}_{n}=U^{-t}J_{n}U^{t}=J_{n}+{c\ov3}t\dlt_{n,0},\\
		&\tilde{L}_{n}=U^{-t} L_{n} U^{t}=L_{n}+tJ_{n}+{c\ov 6}t^{2}\dlt_{n,0}
		\label{1.Sflow},
	\end{aligned}
\end{equation}
known as spectral flow, is very important in the superstring compactification construction.
It allows to relate various representations of $N=2$ Virasoro superalgebra. In particular for $t\in \mathbb{Z}+{1\ov 2}$ the spectral flow interpolates between $NS$  and $R$  representations, while for $t\in\mathbb{Z}$ it takes $NS$ to $NS$ and $R$ to $R$.  

To construct superstring physical states in orbifold compactification, it is important to express the spectral flow operator $U^{t}$ in terms of the free scalar field $\varphi(z)$. It is really possible because the second and third commutation relations from (\ref{1.N2Vir}) mean that $J_{n}$ are the Fourier modes of the free bosonic $U(1)$-current $J(z)$, 
which can be written as
\begin{equation}
J(z)=\im\sqrt{c\ov 3}{\d\varphi\ov \d z},
\label{1.Jboson}
\end{equation}
where $\varphi(z)$ is free massles scalar field.
Then the relations (\ref{1.Sflow}) imply
\begin{equation}
U^{t}=\exp{(\im t\sqrt{c\ov 3}\varphi(z))}.
\label{1.Uboson}
\end{equation}

In Gepner's approach \cite{Gep} the compact sector $M_{int}$ is some orbifold 
of  the  product of $N=2$ minimal models $M_{\bk}$ with the total  central charge $c_{int}=9$, obtained as a result of factorization over some so-called admissible
group $G_{adm}$
\ber
M_{int}=M_{\bk}/G_{adm}=(\prod_{i=1}^{r}M_{k_{i}})/G_{adm}.
\enr
The central charge of the Minimal model $M_{k_{i}}$ is 
$c_{i}=\frac{3k_{i}}{k_{i}+2}$, where $k_{i}=0,1,2,...$.

For these values of $c_{i}$, the model $M_{k_{i}}$ is unitary and consists of a finite number of integrable representations of the  $N=2$ Virasoro superalgebra.
Since the total central charge of the product of $N=2$ minimal models
is equal $9$, this imposes the following conditions
on the set $k_i$
\be
\sum _{i} \frac{ k_{i}}{k_{i}+2}=3.
\ee
If the number of minimal models $r$ is odd, this means that
\be 
\sum _{i} \frac{1}{k_{i}+2}\in Z.
\ee
The representations are generated from the highest primary states $\Phi^{NS,R}_{l,q}$, where $l=0,1,...,k$, $q=-l,-l+2,...,l$.
The conformal dimension and $U(1)$ charge of the state $\Phi^{NS}_{l,q}$ are
\begin{equation}
\Delta_{l,q}=\frac{l(l+2)-q^{2}}{4(k+2)}, \ \ Q_{l,q}=\frac{q}{k+2}.
\label{1.DeltQ}
\end{equation}
The dimension and charge of the primary state $\Phi^{R}_{l,q}$ in $R$ sector are
\begin{equation}
\Delta^{R}_{l,q}=\frac{l(l+2)-(q-1)^{2}}{4(k+2)}+\frac{1}{8},\quad
Q^{R}_{l,q}=\frac{q-1}{k+2}+\frac{1}{2}.
\label{1.DeltQR}
\end{equation}
In the right-moving sector of the Minimal model we have the similar set of integrable representations.

The conformal symmetry of the model $M_{\bk}$ is given by the diagonal $N=(2,2)$ Virasoro algebra acting in the tensor product of individual minimal models $M_{k_{i}}$, whose generators are
\begin{equation}
L_{int,n}=\sum_{i}L_{(i),n}, \ J_{int,n}=\sum_{i}J_{(i),n},
\
G^{\pm}_{int,r}=\sum_{i}G^{\pm}_{(i),r}.
\label{1.DiagVir}
\end{equation}
The algebra action (\ref{1.DiagVir}) is correctly defined on the product of only $NS$-representations or on the product of only $R$-representations of minimal models $M_{k_{i}}$ \cite{Gep}.
The primary states of the  diagonal $N=2$ Virasoro algebra in $NS$ and $R$ sectors  are also products of the primary states from each minimal factor:
\be
\begin{aligned}
	&\Phi^{NS}_{\vec{l},\vec{q}}(z)=\prod_{i}\Phi^{NS}_{l_{i},q_{i}}(z),\\	
	&\Phi^{R}_{\vl,\vq}(z)=\prod_{i}\Phi^{R}_{l_{i},q_{i}}(z).
	\label{1.Prim}
\end{aligned}
\ee
The left-moving  vertex operators of the string in the compact factor can be represented as
\ber
V_{int}(z)=P_{int}(T_{i}, J_{i}, G^{\pm}_{i})\Phi^{NS,R}_{\vec{l},\vec{q}}(z),
\enr
where $P_{int}$ denotes the descendants contributions  of the product model $M_{\bk}$.

As each factor in the product model $M_{\bk}$ has 
$\mathbb{Z}_{k_{i}+2}$ group of symmetries \cite {ZF,GQ, Gep}
 the group of discrete symmetries of 
$M_{\bk}$ containes the group $G_{tot}=\mathbb{Z}_{k_{1}+2}\times...\times\mathbb{Z}_{k_{r}+2}$.

Below we will consider orbifolds of the form $M_{\bk}/G_{adm}$,
where an admissible group $G_{adm}$ by definition is any subgroup of
 $G_{tot}$, whose generators include the distinguished element
$\bet^{0}=(1,1,1,...,1)$, as well as  elements
${\bw}=\{w_1,w_2,...,w_r\}$, which satisfy the following relations
\cite{Gep1}
\be
	\sum_{i} \frac{w_i}{(k_i+2)}\in\mathbb{Z}. 
	\label{1.Gadm}
\ee 
In the $M_{\bk}/G_{adm}$ model, the group $G_{adm}$ occurs twice as it will be expained later.
First, it generates twisted sectors by the products of spectral flow vertices for each minimal model $M_{k_i}$
\begin{equation}
U^{\bw}(z)=\prod_{i}U_{i}^{w_{i}}(z), \ 
U_{i}=\exp{[\im\sqrt{\frac{k_{i}}{k_{i}+2}}\phi_{i}(z)]},
\label{1.UGadm}
\end{equation}
where we introduced the fields $\phi_{i}(z)$, $i=1,...,r$ for the bosonization of left $U(1)$ currents and spectral flow operators of individual submodels according to (\ref{1.Jboson}), (\ref{1.Uboson}).
The second implementation of the group $G_{adm}$ is given by the product of the $U(1)_i$ currents of  $M_{k_i}$
\begin{equation}
\hat{g}^{\bw}=\prod_{i}g_{i}^{w_{i}}=\exp{[\im\pi\sum_{i}w_{i}(J_{i,0}+\bar{J}_{i,0})]}.
\label{1.JGadm}
\end{equation}
When constructing orbifolds of Gepner models, the first implementation
$G_{adm}$ is used to generate a complete set of vertex operators in the twisted sector, consistent with the spacetime symmetry.
As for the second implementation, it is used to select from this set those vertex operators that satisfy mutual locality.

It is noteworthy that the group $G_{adm}$ is naturally accompanied by the mirror
group $G^{*}_{adm}$ \cite{BH}, see also \cite{BE}, which is defined as
\begin{equation}
G^{*}_{adm}=\lbrc {\bw}^{*}\in G_{tot}, \sum_{i=1}^{r}\frac{w_{i}w^{*}_{i}}{k_{i}+2}\in\mathbb{Z}\quad\text{for  any}\quad {\bw}\in G_{adm}  \rbrc.
\label{1.DualGadm}
\end{equation}
The group $G^{*}_{adm}$, as it can be seen from its definition,  includes the generator  $\bet^{0}=(1,1,1,...,1)$.
The group $G^{*}_{adm}$ defines another $N=(2,2)$ superconformal model 
$M_{\bk}/G^{*}_{adm}$, which is the mirror model \cite{DG,LVW} of $M_{ \bk}/G_{adm}$, in which the roles of the groups $G_{adm}$ and $G^{*}_{adm}$ are swapped \cite{ COGP} (see also \cite{P}). 

The  physical states of our string theory    obey the constraints 
given by  the Super Virasoro algebra whose currents are
\begin{equation}
T_{mat}(z)\equiv T_{st}(z)+T_{int}(z),
\ G_{mat}(z)\equiv G_{st}(z)+G^{+}_{int}(z)+G^{-}_{int}(z)
\end{equation}
We will use the BRST approach, in which the physical states
are defined as BRST-invariant fields.
The BRST charge in the left-moving sector includes, along with
currents $T_{mat}(z)$ and $G_{mat}(z)$ also
two fermion ghosts $b(z),c(z)$ and two bosonic ghosts $\bet(z), \gm(z)$.
The explicit expression for the left-moving $BRST$ charge is 
\begin{equation}
Q_{BRST}=\oint dz (cT_{mat}+\gm G_{mat}+\fr12(cT_{gh}+\gm G_{gh})).
\label{1.BRST}
\end{equation}
Left-moving $bc-\beta\gamma$ ghost fields have the OPE's
\begin{equation}
\bet(z)\gm(0)=-z^{-1}+..., \ \ b(z)c(0)=z^{-1}+....
\label{1.ghope}
\end{equation}
$N=1$ superconformal symmetry in the ghost factor is given by the currents
\begin{equation}
\begin{aligned}
&T_{gh}=-\d bc-2b\d c-\frac{1}{2}\d\bet\gm-\frac{3}{2}\bet\d\gm,
\nmb
&G_{gh}=\d\bet c+\frac{3}{2}\bet\d c-2 b\gm.
\label{1.Virgh}
\end{aligned}
\end{equation}
$\bet-\gm$ space of states is characterized by the vacua $V_{q}(z)$. They are parameterized by the  picture number $q$ \cite{BLT}.
The vacuum $V_{q}$ is determined by
\begin{equation}
\bet(z)V_{q}(0)\sim O(z^{q}),
\
\gm(z)V_{q}(0)\sim O(z^{-q}),
\label{1.picvac}
\end{equation}
and can be written as the free field exponential:
\begin{equation}
V_{q}(z)=\exp{[q\phi(z)]},\quad
\phi(z)\phi(0)=-\log(z)+....
\end{equation}
The left-moving ghost vertex operators can be represented as
\begin{equation}
V_{gh}(z)=P_{gh}(\bet,\gm,b,c)\exp{[q\phi]}(z),
\label{1.Vgh}
\end{equation}
where $P_{gh}(\bet,\gm,b,c)(z)$ is a polynomial of $\bet,\gm,b,c$ and their derivatives.
The right-moving physical states are determined similarly.

For the BRST charge to be well defined, the integrand in (\ref{1.BRST}) must be periodic around zero. Therefore, the fields $\bet,\gm$ must have the same monodromy as the fermionic degrees of freedom of the matter.
The ghost $V_{gh}$ and the matter components $V_{st}V_{int}$ of the left-moving vertex  operator at the point $z = 0$ must be in the same sector. 
Therefore, q must be an integer in the $NS$-sector and half-integer in the $R$-sector.
The total vertices ${\cal{V}}(z,\bar{z})$ are the products of the left-moving and the right-moving vertices of ghost, space-time and compact factors
\begin{equation}
{\cal{V}}(z,\bar{z})=V_{gh}V_{st}V_{int}(z)\bar{V}_{gh}\bar{V}_{st}\bar{V}_{int}(\bar{z}).
\label{1.Vert}
\end{equation}
The left and the right factors of vertex operators can independently belong to sectors $NS$ or $R$. So we have $(NS,NS)$, $(NS,R)$, $(R,NS)$, $(R,R)$ types of the vertex operators.

\section{Massless vertex operators and space-time supersymmetry}
\label{sec:3}
Imposing the BRST-invariance requirement, we find the following  massless left vertices in the sector $NS$ with the picture number $(-1)$ and in the sector $R$ with the picture number $(-\fr12)$.
We get the left-moving  vertex  of massless vector boson 
in $(-1)$ picture
\begin{equation}
\exp(-\phi(z)) \xi_{\mu}(p)\psi^{\mu}(z) \exp(\im p_{\mu}X^{\mu})(z), 
\quad
\xi_{\mu}(p)p^{\mu}=0.
\label{2.Vect}
\end{equation}
In the picture $(-\fr12)$ we find four pairs of BRST-invariant currents, which are the vertex operators of massless fermions in the limit of momenta equal to zero
\begin{equation}
\begin{aligned}
&S^{\pm}_{\sigma}(z)=\exp(-\fr12\phi(z) + \im {\sigma}^{a}H_{a}(z) 
\pm\frac{\im}{2}\sum_{i}\frac{k_{i}\phi_{i}(z)}{\sqrt{k_{i}(k_{i}+2)}}),
\sigma^{a}=\pm\fr12,  \ \sum_{a=1}^{2}\sigma^{a}=\pm 1,\\
&S^{\pm}_{\dot{\sigma}}(z)=\exp(-\fr12\phi(z) + \im \dot{{\sigma}}^{a}H_{a}(z)  \pm\frac{\im}{2}\sum_{i}\frac{k_{i}\phi_{i}(z)}{\sqrt{k_{i}(k_{i}+2)}}), 
\dot{ \sigma}^{a}=\pm\fr12,  \ \sum_{a=1}^{2}\dot{\sigma}^{a}=0.
\end{aligned}
\end{equation}
These currents are nothing else but the product of the Ramond vacua of the three sub-theories of the string model under consideration.
In particular, the exponential factors of the internal model $M_{\bk}$ are products of the spectral flow operators for each minimal model $M_{k_{i}}$.
The results in the right sector will be similar. 

It is easy to verify that each of the four currents 
$S^{+}_{\sigma}$, $S^{+}_{\sigma}$,  $S^{+}_{\dot{\sigma}}$ and $S^{-}_{\dot{\sigma}}$,  belongs to the two-dimensional representation 
of the $SO(1,3)$ Lie algebra of space-time symmetry, namely, each of them  is  a  Weyl spinor.
Two currents $S^{+}_{\sigma}$ and $S^{-}_{\sigma}$ have the same chirality, say left-handed,  and the other two, $S^{+}_{\dot{\sigma}}$ and $S^{-}_{\dot{\sigma}}$, have opposite chirality, respectively right-handed.
For our construction, the mutual locality of these currents is important.
A simple check shows that the current  $S^{+}_{\sigma}$ is mutually local with $S^{-}_{\dot{\sigma}}$  and  $S^{-}_{\sigma}$ is mutually 
local with $S^{+}_{\dot{\sigma}}$. 

The four supercharges, defined by the integration of these  currents, define, together with generators of  Poincaré algebra operators, two $N=1$ super-Poincaré algebras of opposite chiralities.
The currents $S^{+}_{\sigma}$ and $S^{-}_{\dot{\sigma}}$ define
super generators  of  a $N=1$ super Poincaré for the case 
of left chirality as
\begin{equation}
\begin{aligned}
&{\cal{Q}}^{(-\fr12)}_{\sigma}= \oint S^{+}_{\sigma}  dz 
=\oint dz \exp{[-\fr12\phi+\im {\sigma}^{a}H_{a}+\frac{\im}{2}
	\sum_{i}\frac{k_{i}\phi_{i}}{\sqrt{k_{i}(k_{i}+2)}}]},\\ 
&{\cal{Q}}^{(-\fr12)}_{\dot{\sigma}}= \oint S^{-}_{\dot{\sigma}} dz
=\oint dz \exp{[-\fr12\phi+\im \dot{\sigma}^{a}H_{a}-\frac{\im}{2}
	\sum_{i}\frac{k_{i}\phi_{i}}{\sqrt{k_{i}(k_{i}+2)}}]}. 
\end{aligned}
\label{2.LSUSYa}
\end{equation}

Instead of the mutually local pair $S^{+}_{\sigma}$ and
 $S^{-}_{\dot{\sigma}}$ we can take another pair $S^{-}_{\sigma}$ and $S^{+}_{\dot{\sigma}}$.
As a result of either of these two different options, we obtain  $d=4$ $N=1$ spacetime supersymmetry. For each of these two options, we have to leave from the BRST-invariant fields only those fields that are mutually local with respect to the chosen supercharges.

We can make such choice in both   left-moving and right-moving sectors.
Thus, we obtain string models $IIB$ or $IIA$ according to
whether the choice of super-charges in the left-moving and right-moving sectors is the same or not.
\section{Space-time supersymmetry and GSO equations} \label{sec:4}
In general, the vertex operators  are not mutually local with the currents of supersymmetry operators.

In order to get SUSY algebra representation we select only the vertices which are mutually local with SUSY currents. 
For the $IIB$ case this requirement gives the GSO equations
\begin{equation}
\begin{aligned}
&q+\sum_{a}\lm_{a}+Q_{int}\in 2\mathbb{Z},\\
&\bar{q}+\sum_{a}\bar{\lm}_{a}+\bar{Q}_{int}\in 2\mathbb{Z},
\end{aligned}
\label{3.GSOB}
\end{equation}
where $Q_{int}$, $\bar{Q}_{int}$ are the $U(1)$ charges of the states from the internal sector. 

For the type $IIA$  the second equation changes
\begin{equation}
\begin{aligned}
&q+\sum_{a}\lm_{a}+Q_{int}\in 2\mathbb{Z},\\
&\bar{q}+\sum_{a}\bar{\lm}_{a}-\bar{Q}_{int}\in 2\mathbb{Z},
\end{aligned}
\label{3.GSOA}
\end{equation}
\section{Construction of physical vertices.} \label{sec:5}
We now move on to the construction of complete physical vertex operators, which are products of left- and right-moving factors.
Let's start doing this for theory $IIB$.
First, we construct the corresponding set of the vertex operators
type $(NS,NS)$, which are consistent with the action of space-time supersymmetry and satisfy the requirements of  BRST invariance and 
mutual locality.

To analyze the problem of mutual locality and supersymmetry, we consider only the part of the $(NS,NS)$ vertex operators (\ref{1.Vert}) which is essential for this issue, i.e. the operators  which do not include the descendants because in $(NS,NS)$ sector the descendants can not spoil mutual locality:                             
\begin{equation}
{\cal{V}}^{(q,\lm)(\bar{q},\bar{\lm})}_{\bl,\bar{\bt},\bw}
(z,\bar{z}) =
\exp{[q\phi+\im\lm^{a}H_{a}]}(z)
\exp{[\bar{q}\bar{\phi}+\im\bar{\lm}^{a}\bar{H}_{a}]}(\bar{z})
\Psi_{\bl,{\bar{\bt}},{\bw}}(z,\bar{z}).
\label{4.NSNS}
\end{equation}
Here the  factor $\Psi_{\bl,{\bar{\bt}},{\bw}}(z,\bar{z})$ 
from the compact sector is given by application of the left-moving spectral flow operator $U^{\bw}$ to the  state  $ \Phi_{\bl,\bar{\bt}}(z) \bar{\Phi}_{\bl,\bar{\bt}}(\bar{z})$
in the product model $M_{\bk}$.  This construction  and the corresponding explicit expressions are given in (\ref{A.Primo})-(\ref{A.Primo2}) of the Appendix.

The vectors $\bw$ label the orbifolds twisted sectors, which for the orbifolds we consider are $r$-dimensional vectors 
${\bw}=(w_{1},...,w_{r})$, where $w_{i}$ are integers and 
$\mod (\ k_{i}+2)$ defined. Thus, when ${\bw=}0$ the field $\Psi_{\bl,{\bar{\bt}},{\bw}}(z,\bar{z})$ is the diagonal product of the left-moving and the right-moving primary states 
$\Phi_{\bl,\bar{\bt}}(z) \bar{\Phi}_{\bl,\bar{\bt}}(\bar{z})$ from the product model $M_{\bk}$. When ${\bw}\neq 0$ the field $\Psi_{\bl,{\bar{\bt}},{\bw}}(z,\bar{z})$ is no longer the diagonal product (but this field is still diagonal along the indices $\bl$), but spectral flow twisted product.

The reason why the twisted vertices (\ref{4.NSNS}) are necessary in the superstring construction is the space-time supersymmetry requirement. Indeed, if one applies type $IIB$ SUSY operators  twice to the vertex (\ref{4.NSNS}) with ${\bw}=0$
we obtain  $(NS,NS)$ vertices (\ref{4.NSNS}), where ${\bw}=(1,1,...,1)$.
Therefore, the minimum set of $(NS,NS)$ vertices which is consistent with the SUSY action must be of the form (\ref{4.NSNS}), where ${\bw}=(n,n,...,n )=n{\bet}^{0}$.

To include more general orbifolds in the construction, it is necessary to consider twist vectors $\bw$ of a more general form, but such that the vertices (\ref{4.NSNS}) satisfy the GSO equations \eqref{3.GSOB}.  

Using the expressions (\ref{A.Primo})-(\ref{A.Primo2}) we find that GSO equations for the vertices (\ref{4.NSNS}) takes the form
\begin{equation}
\begin{aligned}
&q+\sum_{a}\lm_{a}+\sum_{i=1}^{r}\frac{q'_{i}-2w_{i}}{k_{i}+2}+Q_{des}\in 2\mathbb{Z},
\\
&\bar{q}+\sum_{a}\bar{\lm}_{a}+\sum_{i=1}^{r}\frac{\bar{q}_{i}}{k_{i}+2}+\bar{Q}_{des}
\in 2\mathbb{Z},
\label{4.GSOB}
\end{aligned}
\end{equation}
where $\bar{q}_{i}=l_{i}-2\bar{t}_{i}$, $q'_{i}=\bar{q}_{i}$ if $0\leq \bar{t}_{i}+w_{i}\leq l_{i}$, or $q'_{i}=k_{i}+2+\bar{q}_{i}$ if $l_{i}+1\leq \bar{t}_{i}+w_{i}\leq k_{i}+1$.
$Q_{des},\bar{Q}_{des}$ are the charges contributions of the descendants, 
which were omitted in (\ref{4.NSNS}). 

We see that the requirement 
\begin{equation}
\sum_{i=1}^{r}\frac{w_{i}}{k_{i}+2}\in\mathbb{Z}
\label{4.Gadm}
\end{equation}
is consistent with the GSO equations. We will see soon that this condition follows from the mutual locality requirement.
We denote the complete set of solutions (\ref{4.Gadm}), including the vector $\bet^{0}=(1,1,...,1)$, as the maximal admissible group $G^ {max}_{adm }$.
This set forms an additive group due to OPE requirements.
Let us fix some admissible subgroup  $G_{adm}=\lbrc {\bw}=\sum_{a=0}^{K-1}w_{a}\bet^{a}, \ \bet^{0}=(1,1,...,1), \ w_{a}\in\mathbb{Z}\rbrc\subset G^{max}_{adm}$, where the vectors $\bet^{a}$ are generators of $G_{adm}$.
We will use this group to construct $IIB$ superstring compactification on the orbifold $M_{\bk}/G_{adm}$.

We want to consider the question of mutual locality of 
$(NS,NS)$ vertex operators (\ref{4.NSNS}).	 
Then for the pair of vertex operators 
${\cal{V}}^{(q_{1},\lm_{1})(\bar{q}_{1},\bar{\lm}_{1})}_{\bl_{1},\bar{\bt}_{1},\bw_{1}}(z,\bar{z})$ and
${\cal{V}}^{(q_{2},\lm_{2})(\bar{q}_{2},\bar{\lm}_{2})}_{\bl_{2},{\bar{\bt}}_{2},{\bw}_{2}}(0)$ we have to calculate the monodromy when the first vertex goes around the second one.
Since they are in the $(NS,NS)$ sector, the nontrivial monodromy phase for this pair arises only from compact factors. As a result, the mutual locality equation, as shown in the Appendix, has the form
\begin{equation}
\sum_{i=1}^{r}\frac{w_{1i}(\bar{q}_{2i}-w_{2i})+w_{2i}(\bar{q}_{1i}-w_{1i})}{k_{i}+2}\in\mathbb{Z}.
\label{4.mloc}
\end{equation}
These equations select for each pair of vectors ${\bw}_{1,2}$ the set of $\bar{\bq}_{1,2}$ which give mutually local vertices. 

As one can show, from these equations it follows that the charges $\bar{q}_i$ of all vertices
$\Psi^{(q,\lm)(\bar{q},\bar{\lm})}_{\bl,\bar{\bt},{\bw}}$ 
deformed by some  twist ${\bw} \in G_{adm}$ must satisfy the equation 
\begin{equation}
\sum_{i=1}^{r}\frac{\bet^a_{i}(\bar{q}_{i}-w_{i})}{k_{i}+2}\in\mathbb{Z}
\label{4.BHK}
\end{equation}
for all generators  $\bet^a \in G_{adm}$. 
These equations mean that the set of $U(1)$ charges of the mutually local vertex operators are such that they are invariant w.r.t. the defined above 
operators (\ref{1.JGadm})
\begin{equation}
\hat{g}^{\bw}=\exp{[\im\pi\sum_{i}w_{i}(J_{i,0}+\bar{J}_{i,0})]}, 
\end{equation}
that is, such vertex operators  (\ref{4.NSNS}) are $G_{adm}$-invariant.

Moreover, from the last equations it follows that the vectors
$\bw^{*}\equiv \bar{\bq}-\bw$ form another admissible group
$G^{*}_{adm}\subset G^{max}_{adm}$ \cite{P}.
This group is Berglund-Krawitz-Hubsh dual group for the Fermat type potential \cite{BH}. 
It defines the mirror orbifold model $M_{\bk}/G^{*}_{adm}$ \cite{P}. 

This  fact allows to represent the left
$\bq=\bar{\bq}-2\bw$ and the right  $\bar{\bq}$  vectors 
labeling the mutually local vertices of the twisted sector 
$\bw$ of the model $M_{\bk}/G_{adm}$  in the mirror symmetric form 
\begin{equation}
\bq={-\bw+\bw^{*}},
\quad
{\bar{\bq}}={\bw+\bw^{*}},
\label{4.mloc2}
\end{equation}
where ${\bw}\in G_{adm}, {\bw}^{*}\in G^{*}_{adm}$.

Note that the requirement (\ref{4.Gadm}) we imposed earlier for the group $G_{adm}$
now follows from (\ref{4.BHK}) but for the group $G^{*}_{adm}$, since $G_{adm}$ contains $\bet^{0}$.
Hence, (\ref{4.Gadm}) also follows from (\ref{4.BHK}) for the group $G_{adm}$ because of the mirror symmetry.

In order to get the total $BRST$ invariant vertex from (\ref{4.NSNS}) one needs to recover the descendant contributions in an appropriate way. 
\section{Supersymmetry and mutual locality in $(R,NS)$, $(NS,R)$ and $(R,R)$ sectors.}
\label{sec:6}
In this section we generate physical $(R,NS)$, $(NS,R)$, $(R,R)$ vertices from the $(NS,NS)$ physical vertices constructed in the previous section by applying the type $IIB$ supercharges. 
We then show the mutual locality of these vertices both with themselves and with the $(NS,NS)$ vertices we started with.

Let's consider the physical vertex operators $(NS,NS)$ constructed in the previous section and apply the left supercharges to them.
 We obtain physical $(R,NS)$ vertex, because the super-charges commute with $Q_{BRST}$. Modulo the descendants factor the result of the supersymmetry action can be written in the form
\begin{equation}
\tld{{\cal{V}}}^{(q,\lm)(\bar{q},\bar{\lm})}_{\bl,\bar{\bt},\bw}(z,\bar{z})=\exp{[(q-\fr12)\phi+\im(\lm^{a}+\sgm^{a})H_{a}]}(z)
\exp{[\bar{q}\bar{\phi}+\im\bar{\lm}^{a}\bar{H}_{a}]}(\bar{z})
U^{\fr12\bet^{0}}\Psi_{\bl,\bar{\bt},\bw}(z,\bar{z}),
\label{5.RNS}
\end{equation}

Similarly, we obtain $(NS,R)$ physical vertex by applying type $IIB$ right-moving super-charges to some other physical $(NS,NS)$ vertex constructed in the previous section. Modulo the descendants factor the result of the supersymmetry action is given by
\ber
\tld{\tld{{\cal{V}}}}^{(q,\lm)(\bar{q},\bar{\lm})}_{\bl,\bar{\bt},\bw}(z,\bar{z})=\exp{[q\phi+\im\lm^{a}H_{a}]}(z)
\exp{[(\bar{q}-\fr12)\bar{\phi}+\im(\bar{\lm}^{a}+\bar{\sgm}^{a})\bar{H}_{a}]}(\bar{z})
\bar{U}^{\fr12\bet^{0}}\Psi_{{\bl},\bar{\bt},\bw}(z,\bar{z}).
\label{5.NSR}
\enr

Finally, we obtain $(R,R)$ sector vertex by applying left-moving and right-moving type $IIB$ super-charges to the physical $(NS,NS)$ vertex from the previous section.  
Again, modulo the descendants factor the result is given by
\ber
\begin{aligned}
&\tld{\tld{\tld{{\cal{V}}}}}^{(q,\lm)(\bar{q},\bar{\lm})}_{\bl,\bar{\bt},\bw}(z,\bar{z})=
\\
&\exp{[(q-\fr12)\phi+\im(\lm^{a}+\sgm^{a})H_{a}]}(z)
\exp{[(\bar{q}-\fr12)\bar{\phi}+\im(\bar{\lm}^{a}+\bar{\sgm}^{a})\bar{H}_{a}]}(\bar{z})
U^{\fr12\bet^{0}}\bar{U}^{\fr12\bet^{0}}\Psi_{{\bl},\bar{\bt},\bw}(z,\bar{z}).
\label{5.RR}
\end{aligned}
\enr

In order to prove mutual locality of the superpartners, arising in this way, we need to calculate the monodromy phases between 
each of these vertices and an arbitrary $(NS,NS)$ vertex constructed in the previous section as well as between these vertices themselves. 

Carrying out the calculation of phases one can ignore at first step the descendants contributions and calculate the phases between the vertices in the form (\ref{5.NSR})-(\ref{5.RR}). Then we recover descendants contributions. This is given in the Appendix. The result shows that the phases are trivial because the original $(NS,NS)$ vertices where mutually local and obeyed $GSO$ equations (\ref{4.GSOB}).


Thus, we have constructed the physical state space of type $IIB$ superstring compactified on the orbifold $M_{\bk}/G_{adm}$. 
Similarly, one can construct the physical state space of a type $IIA$ superstring.

\section{Modular invariance}\label{sec:7}
The equations (\ref{4.mloc2}) can be rewritten in the form
\ber
{\bq+\bar{\bq}}\in G^{*}_{adm}, \ {\bar{\bq}-\bq}\in G_{adm}.
\label{6.GP}
\enr
We come thereby to the constraints for the modular invariant partition function of the Gepner model orbifold obtained in \cite{GrPl}. Therefore, the partition function of the superstring calculated in our approach will also be modular invariant.
\section{Conclusion}
\label{sec:8}
We are grateful to D. Gepner, A. Litvinov and A. Marshakov for useful discussions. The work is supported by RSF grant 23-12-00333.
\section{Appendix}\label{sec:9}
{\bf A.1.} For the minimal model $M_{k}$ spectral flow can be used to construct all primary fields starting from chiral or anti-chiral primary fields \cite{BBP}, \cite{P}. 
Here we briefly discuss the construction, following mainly to the papers 
\cite{BBP}, \cite{P}, where more details can be found. 

First of all recall the definition of primary state in $NS$ sector
\begin{equation}
	\begin{aligned}
				&L_{n}\Phi^{NS}_{\Delta,Q}=0, \ J_{n}\Phi^{NS}_{\Delta,Q}=0, \ n\geq 1,
			\
			G^{\pm}_{r}\Phi^{NS}_{\Delta,Q}=0, \ r\geq {1\ov 2},
			\\
			&L_{0}\Phi^{NS}_{\Delta,Q}=\Delta \Phi^{NS}_{\Delta,Q},
			\quad
			J_{0}\Phi^{NS}_{\Delta,Q}=Q\Phi^{NS}_{\Delta,Q},
			\label{A.NSPrim}
		\end{aligned}
\end{equation} 
 
Chiral primary state is subject to an additional constraint
\ber
G^{+}_{-{1\ov 2}}\Phi^{NS}_{\Delta,Q}=0.
\label{A.CPrim}
\enr
It leads to the relation between conformal dimension and $U(1)$ charge of the chiral primary state: $Q=2\Delta$. The anti-chiral primary is defined similarly
\ber
G^{-}_{-{1\ov 2}}\Phi^{NS}_{\Delta,Q}=0
\label{A.APrim}
\enr 
and for this field  $Q=-2\Delta$.

For the Minimal model $M_{k}$, the chiral and anti-chiral primary states are the primaries with $q=l$ and $q=-l$ correspodingly:
\ber
\Phi^{c}_{l}(z)\equiv\Phi^{NS}_{l,l}(z), \ \Phi^{a}_{l}(z)\equiv\Phi^{NS}_{l,-l}(z).
\label{A.MCAPrim}
\enr
 
All other primary states in the Minimal model can be constructed starting for example from chiral primaries. Namely, for the spectral flow parameter $0\leq t\leq l$, the state 
\ber
\Phi_{l,t}=(UG^{-}_{-{1\ov 2}})^{t}\Phi^{c}_{l}
\label{A.MCPrim1}
\enr
gives spectral flow realization of the primary state $\Phi^{NS}_{l,q}$, where  $q=l-2t$.
For the spectral flow parameter $l+1\leq t\leq k+1$, the state
\ber
\Phi_{l,t}=(UG^{-}_{-{1\ov 2}})^{t-l-1}U(UG^{-}_{-{1\ov 2}})^{l}\Phi^{c}_{l}
\label{A.MCPrim2}
\enr
gives the spectral flow realization of the primary state $\Phi^{NS}_{\tld{l},\tld{q}}$, where $\tld{l}=k-l$ and $\tld{q}=k+2+l-2t$.

Notice that in the expressions (\ref{A.MCPrim1}),(\ref{A.MCPrim2}) we use spectral flow parameter $t$ to label primary states. 

It is clear how to extend the spectral flow construction above for the primary states in the product model $M_{\bk}$. Moreover, the spectral flow construction is very helpfull for the construction of the $(NS,NS)$ primary fields of the orbifold $M_{\bk}/G_{adm}$ in the twisted sectors. Indeed, for the twisted sector ${\bw} \in G_{adm}$ the twisted primary field is given by
\ber
\Psi_{\bl,\bar{\bt},\bw}(z,\bar{z})=
\Phi_{\bl,\bar{\bt}+\bw}(z)\bar{\Phi}_{\bl,\bar{\bt}}(\bar{z}),
\label{A.Primo}
\enr
where
\ber
\bar{\Phi}_{\bl,\bar{\bt}}(\bar{z})=\prod_{i}\bar{\Phi}_{l_{i},\bar{t}_{i}}(\bar{z}),\
\bar{\Phi}_{l_{i},\bar{t}_{i}}(\bar{z})=(\bar{U}\bar{G}^{-}_{-{1\ov 2}})_{i}^{\bar{t}_{i}}\bar{\Phi}^{c}_{l_{i}}(\bar{z}), \ 0\leq \bar{t}_{i}\leq l_{i},
\label{A.Primo1}
\enr
and
\ber
\Phi_{\bl,\bar{\bt}+\bw}(z)=
\prod_{i=1}^{5}\Phi_{l_{i},\bar{t}_{i}+w_{i}}(z),
\enr
where
\ber
\begin{aligned}
	\Phi_{l_{i},\bar{t}_{i}+w_{i}}(z)=\begin{cases}(UG^{-}_{-{1\ov 2}})_{i}^{\bar{t}_{i}+w_{i}}\Phi^{c}_{l_{i}}(z), 
		\qquad \text{if}\quad \ 0\leq \bar{t}_{i}+w_{i}\leq l_{i}, \\\\
		(UG^{-}_{-{1\ov 2}})_{i}^{\bar{t}_{i}+w_{i}-l_{i}-1}
		U_{i}(UG^{-}_{-{1\ov 2}})_{i}^{l_{i}}\Phi^{c}_{l_{i}}(z), 
		\quad \text{if}\quad \ l_{i}+1\leq \bar {t}_{i}+w_{i}\leq k_{i}+1.
	\end{cases}
	\label{A.Primo2}
\end{aligned}
\enr

Applying spectral flow operators $U^{\fr12\bet^{0}}=\prod_{i=1}^{r}U_{i}^{\fr12}$ and $\bar{U}^{\fr12\bet^{0}}=\prod_{i=1}^{r}\bar{U}_{i}^{\fr12}$ to the fields (\ref{A.Primo}) independently we can construct primary fields of the orbifold in $(R,NS)$, $(NS,R)$ and $(R,R)$ sectors.

{\bf A.2.} Here we derive mutual locality condition (\ref{4.mloc}). 

To do this, we calculate the monodromy for an arbitrary pair of vertices
 inside sector $(NS,NS)$
\ber
{\cal{V}}^{(q_{1},\lm_{1})(\bar{q}_{1},\bar{\lm}_{1})}_{\bl_{1},\bar{\bt}_{1},\bw_{1}}(z,\bar{z})=\exp{[q_{1}\phi+\im\lm^{a}_{1}H_{a}]}(z)
\exp{[\bar{q}_{1}\bar{\phi}+\im\bar{\lm}^{a}_{1}\bar{H}_{a}]}(\bar{z})
\Psi_{\bl_{1},\bar{\bt}_{1},\bw_1}(z,\bar{z}),
\nmb
{\cal{V}}^{(q_{2},\lm_{2})(\bar{q}_{2},\bar{\lm}_{2})}_{\bl_{2},{\bar{\bt}}_{2},{\bw}_{2}}(z,\bar{z})=
\exp{[q_{2}\phi+\im\lm^{a}_{2}H_{a}]}(z)
\exp{[\bar{q}_{2}\bar{\phi}+\im\bar{\lm}^{a}_{2}\bar{H}_{a}]}(\bar{z})\Psi_{\bl_{2},\bar{\bt}_{2},\bw_{2}}(z,\bar{z}).
\label{A.Loc1}
\enr
 It can be done by calculation of their OPE. 

The fusion 
${\cal{V}}^{(q_{1},\lm_{1})(\bar{q}_{1},\bar{\lm}_{1})}_{\bl_{1},\bar{\bt}_{1},\bw_{1}}(z,\bar{z}){\cal{V}}^{(q_{2},\lm_{2})(\bar{q}_{2},\bar{\lm}_{2})}_{\bl_{2},{\bar{\bt}}_{2},{\bw}_{2}}(0)$
gives the linear combination of vertices 
\begin{equation}
{\cal{V}}^{(q_{3},\lm_{3})(\bar{q}_{3},\bar{\lm}_{3})}_{\bl_{3},\bar{\bt}_{3},\bw_{3}}(0)=\exp{[q_{3}\phi+\im\lm^{a}_{3}H_{a}]}(0)
\exp{[\bar{q}_{3}\bar{\phi}+\im\bar{\lm}^{a}_{3}\bar{H}_{a}]}(0)
\Psi^{{\bl}_{3}}_{\bar{\bt}_{3},\bw_{3}}(0),
\end{equation}
where 
\ber
\begin{aligned}
&q_{3}=q_{1}+q_{2},  \ \bar{q}_{3}=\bar{q}_{1}+\bar{q}_{2}, 
\\
&\lm_{3}=\lm_{1}+\lm_{2}, \ \bar{\lm}_{3}=\bar{\lm_{1}}+\bar{\lm}_{2},\ \bar{\bt}_{3}=\bar{\bt}_{1}+\bar{\bt}_{2},
\\
&\bw_{3}=\bw_{1}+\bw_{2}.
\label{A.conserv}
\end{aligned}
\enr

These vertices (together with the corresponding structure constants) are accompanied by the multipliers
\begin{equation}
z^{-q_{1}q_{2}+\lm_{1}\cdot\lm_{2}+\Delta_{3}-{\Delta_{1}}-{\Delta_{2}}}\bar{z}^{-\bar{q}_{1}\bar{q}_{2}+\bar{\lm}_{1}\cdot\bar{\lm}_{2}+\bar{\Delta}_{3}-\bar{\Delta}_{1}-\bar{\Delta}_{2}}.
\end{equation}
Here, $\Delta_{1,2,3},
\bar{\Delta}_{1,2,3}$ are the conformal dimensions of vertices in the compact sector.

Because we are in $(NS,NS)$ sector nontrivial monodromy phase arises only from compact factors and is given by
\begin{equation}
\exp{[\im 2\pi(\Delta_{3}-\Delta_{1}-\Delta_{2}-\bar{\Delta}_{3}+\bar{\Delta}_{1}+\bar{\Delta}_{2})]}.
\end{equation}

Using (\ref{1.DeltQ}) we find
\begin{equation}
\Delta_{3}-\Delta_{1}-\Delta_{2}-\bar{\Delta}_{3}+\bar{\Delta}_{1}+\bar{\Delta}_{2}=\sum_{i=1}^{r}\frac{1}{4(k_{i}+2)}(q_{1i}^{2}+q_{2i}^{2}-q_{3i}^{2}-\bar{q}_{1i}^{2}-\bar{q}_{2i}^{2}+\bar{q}_{3i}^{2}).
\end{equation}
Taking into account (\ref{A.Primo2}) we find that modulo integer number the expression above can be rewritten as
\ber
\begin{aligned}
&\Delta_{3}-\Delta_{1}-\Delta_{2}-\bar{\Delta}_{3}+\bar{\Delta}_{1}+\bar{\Delta}_{2}=
\\
&\sum_{i=1}^{r}\frac{1}{4(k_{i}+2)}((\bar{q}_{1i}-2w_{1i})^{2}+(\bar{q}_{2i}-2w_{2i})^{2}-(\bar{q}_{3i}-2w_{3i})^{2}-\bar{q}_{1i}^{2}-\bar{q}_{2i}^{2}+\bar{q}_{3i}^{2})=
\\
&\sum_{i=1}^{r}\frac{1}{k_{i}+2}(w_{1i}^{2}-\bar{q}_{1i}w_{1i}+w_{2i}^{2}-\bar{q}_{2i}w_{2i}-w_{3i}^{2}+\bar{q}_{3i}w_{3i}).
\\
\end{aligned}
\enr
Taking into account (\ref{A.conserv}) we find the mutual locality condition
\ber
\sum_{i=1}^{r}\frac{w_{1i}(\bar{q}_{2i}-w_{2i})+w_{2i}(\bar{q}_{1i}-w_{1i})}{k_{i}+2}\in\mathbb{Z}.
\label{A.mloc}
\enr

{\bf A.3.}
 
Here we prove mutual locality of vertices which are $IIB$ SUSY superpartners of mutually local $(NS,NS)$ vertices.

Let us consider the fusion of two vertices $\tld{{\cal{V}}}^{(q_{1},\lm_{1})(\bar{q}_{1},\bar{\lm}_{1})}_{\bl_{1},\bar{\bt}_{1},\bw_{1}}(z,\bar{z}){\cal{V}}^{(q_{2},\lm_{2})(\bar{q}_{2},
\bar{\lm}_{2})}_{\bl_{2},\bar{\bt}_{2},\bw_{2}}(0)$, which were defined in \eqref{5.RNS} and \eqref{4.NSNS}. This fusion gives the linear combination of vertices 
\begin{equation}
\tld{{\cal{V}}}^{(q_{3},\lm_{3})(\bar{q}_{3},\bar{\lm}_{3})}_{\bl_{3},\bar{\bt}_{3},\bw_{3}}(0)=
\exp{[(q_{3}-\fr12)\phi+\im(\lm^{a}_{3}+\sgm^{a})H_{a}]}(0)
\exp{[\bar{q}_{3}\bar{\phi}+\im\bar{\lm}^{a}_{3}\bar{H}_{a}]}(0)
U^{\fr12\bet^{0}}\Psi_{{\bl}_{3},\bar{\bt}_{3},\bw_{3}}(0),
\end{equation}
where $U^{\fr12\bet^{0}}\Psi^{{\bl}_{3}}_{\bar{\bt}_{3},\bw_{3}}(0)$ is $(R,NS)$ primary field and again
\begin{equation}
\begin{aligned}
&q_{3}=q_{1}+q_{2},  \ \bar{q}_{3}=\bar{q}_{1}+\bar{q}_{2}, 
\\
&\lm_{3}=\lm_{1}+\lm_{2}, \ \bar{\lm}_{3}=\bar{\lm_{1}}+\bar{\lm}_{2},\ \bar{\bt}_{3}=\bar{\bt}_{1}+\bar{\bt}_{2},
\\
&\bw_{3}=\bw_{1}+\bw_{2}.
\label{A.conserv1}
\end{aligned}
\end{equation}
These vertices are accompanied by multipliers
\begin{equation}
z^{-(q_{1}-\fr12)q_{2}+(\lm_{1}+\sgm)\cdot\lm_{2}+\Delta_{3}-{\Delta_{1}}-{\Delta_{2}}}\bar{z}^{-\bar{q}_{1}\bar{q}_{2}+\bar{\lm}_{1}\cdot\bar{\lm}_{2}+\bar{\Delta}_{3}-\bar{\Delta}_{1}-\bar{\Delta}_{2}}.
\end{equation}
The monodromy  is given by
\begin{equation}
\exp{[\im 2\pi(\fr12q_{2}+\sgm\cdot\lm_{2}+\Delta_{3}-\Delta_{1}-\Delta_{2}-\bar{\Delta}_{3}+\bar{\Delta}_{1}+\bar{\Delta}_{2})]},
\end{equation}
where $\Delta_{1},\Delta_{3}$ are the dimensions of $R$ sector primary fields and are given by (\ref{1.DeltQR}), the other dimensions are given by (\ref{1.DeltQ}). Using this we obtain
\begin{equation*}
\begin{aligned}
&\Delta_{3}-\Delta_{1}-\Delta_{2}-\bar{\Delta}_{3}+\bar{\Delta}_{1}+\bar{\Delta}_{2}=
\\
&\sum_{i=1}^{r}\frac{1}{4(k_{i}+2)}((q_{1i}-1)^{2}+q_{2i}^{2}-(q_{3i}-1)^{2}-\bar{q}_{1i}^{2}-\bar{q}_{2i}^{2}+\bar{q}_{3i}^{2}).
\end{aligned}
\end{equation*}
Taking into account (\ref{A.Primo2}) we find that modulo integer number this expression takes the form
\ber
\begin{aligned}
&\Delta_{3}-\Delta_{1}-\Delta_{2}-\bar{\Delta}_{3}+\bar{\Delta}_{1}+\bar{\Delta}_{2}=
\\
&\sum_{i=1}^{r}\frac{1}{4(k_{i}+2)}((\bar{q}_{1i}-2w_{1i})^{2}-2(\bar{q}_{1i}-2w_{1i})+(\bar{q}_{2i}-2w_{2i})^{2}-(\bar{q}_{3i}-2w_{3i})^{2}+2(\bar{q}_{3i}-2w_{3i})-
\\
&\bar{q}_{1i}^{2}-\bar{q}_{2i}^{2}+\bar{q}_{3i}^{2}).
\nmb
\end{aligned}
\enr
Using (\ref{A.conserv1}) we obtain
\ber
\begin{aligned}
&\Delta_{3}-\Delta_{1}-\Delta_{2}-\bar{\Delta}_{3}+\bar{\Delta}_{1}+\bar{\Delta}_{2}=
\\
&\sum_{i=1}^{r}(\frac{w_{1i}(\bar{q}_{2i}-w_{2i})+w_{2i}(\bar{q}_{1i}-w_{1i})}{k_{i}+2}+\fr12\frac{\bar{q}_{2i}-2w_{2i}}{k_{i}+2}).
\nmb
\end{aligned}
\enr
Hence, the mutual locality condition is
\begin{equation}
\sum_{i=1}^{r}(\frac{w_{1i}(\bar{q}_{2i}-w_{2i})+w_{2i}(\bar{q}_{1i}-w_{1i})}{k_{i}+2}+\fr12(q_{2}+\sum_{a}\lm^{a}_{2}+\sum_{i=1}^{r}\frac{\bar{q}_{2i}-2w_{2i}}{k_{i}+2})\in\mathbb{Z}.
\end{equation}
The descendants which have been omitted in the vertex ${\cal{V}}^{(q_{2},\lm_{2})(\bar{q}_{2},\bar{\lm}_{2})}_{\bl_{2},\bar{\bt}_{2},\bw_{2}}(0)$ correct the expression above to
\begin{equation}
\sum_{i=1}^{r}\frac{w_{1i}(\bar{q}_{2i}-w_{2i})+w_{2i}(\bar{q}_{1i}-w_{1i})}{k_{i}+2}+\fr12(q_{2}+\sum_{a}\lm^{a}_{2}+\sum_{i=1}^{r}\frac{\bar{q}_{2i}-2w_{2i}}{k_{i}+2}+Q_{2des})\in\mathbb{Z}.
\end{equation}
(It is easy to see that the descendants omitted in vertex $\tld{{\cal{V}}}^{(q_{1},\lm_{1})(\bar{q}_{1},\bar{\lm}_{1})}_{\bl_{1},\bar{\bt}_{1},\bw_{1}}(z,\bar{z})$ do not contribute to the left hand side of this expression.)

Because the initial $(NS,NS)$ vertices where mutually local the first sum in the expression above is integer so we are left with the expression
\begin{equation}
q_{2}+\sum_{a}\lm^{a}_{2}+\sum_{i=1}^{r}\frac{q_{2i}}{k_{i}+2}+Q_{2des}\in 2\mathbb{Z},
\end{equation}
which is nothing else but the first line from (\ref{3.GSOB}).

Similarly, the fusion of $\tld{\tld{{\cal{V}}}}^{(q_{1},\lm_{1})(\bar{q}_{1},\bar{\lm}_{1})}_{\bl_{1},\bar{\bt}_{1},\bw_{1}}(z,\bar{z}){\cal{V}}^{(q_{2},\lm_{2})(\bar{q}_{2},\bar{\lm}_{2})}_{\bl_{2},\bar{\bt}_{2},\bw_{2}}
(0)$, where the first vertex is given by \eqref{5.NSR} gives the linear combination of vertices 
\begin{equation}
\tld{\tld{{\cal{V}}}}^{(q_{3},\lm_{3})(\bar{q}_{3},\bar{\lm}_{3})}_{\bl_{3},\bar{\bt}_{3},\bw_{3}}(0)=
\exp{[q_{3}\phi+\im\lm^{a}_{3}H_{a}]}(0)
\exp{[(\bar{q}_{3}-\fr12)\bar{\phi}+\im(\bar{\lm}^{a}_{3}+\sgm^{a})\bar{H}_{a}]}(0)
\bar{U}^{\fr12\bet^{0}}\Psi_{{\bl}_{3},\bar{\bt}_{3},\bw_{3}}(0),
\end{equation}
where $\bar{U}^{\fr12\bet^{0}}\Psi_{{\bl}_{3},\bar{\bt}_{3},\bw_{3}}(0)$ is $(NS,R)$ sector primary field from the compact part. 

These vertices (together with the corresponding structure constants) are accompanied by multipliers
\begin{equation}
z^{-q_{1}q_{2}+\lm_{1}\cdot\lm_{2}+\Delta_{3}-{\Delta_{1}}-{\Delta_{2}}}\bar{z}^{-(\bar{q}_{1}-\fr12)\bar{q}_{2}+(\bar{\lm}_{1}+\sgm)\cdot\bar{\lm}_{2}+\bar{\Delta}_{3}-\bar{\Delta}_{1}-\bar{\Delta}_{2}}.
\end{equation}
The monodromy is given by
\begin{equation}
\exp{[\im 2\pi(-\fr12\bar{q}_{2}-\sgm\cdot\bar{\lm}_{2}+\Delta_{3}-\Delta_{1}-\Delta_{2}-\bar{\Delta}_{3}+\bar{\Delta}_{1}+\bar{\Delta}_{2})]},
\end{equation}
where $\bar{\Delta}_{1},\bar{\Delta}_{3}$ are the dimensions of $R$ sector primary fields and are given by (\ref{1.DeltQR}), the other dimensions are given by (\ref{1.DeltQ}). Using this we obtain
\be
\begin{aligned}
&\Delta_{3}-\Delta_{1}-\Delta_{2}-\bar{\Delta}_{3}+\bar{\Delta}_{1}+\bar{\Delta}_{2}=
\nmb
&\sum_{i=1}^{r}\frac{1}{4(k_{i}+2)}(q_{1i}^{2}+q_{2i}^{2}-q_{3i}^{2}-(\bar{q}_{1i}-1)^{2}-\bar{q}_{2i}^{2}+(\bar{q}_{3i}-1)^{2}).
\nmb
\end{aligned}
\en
Taking into account (\ref{A.Primo2}) we find that modulo integer number this expression takes the form
\be
\begin{aligned}
&\Delta_{3}-\Delta_{1}-\Delta_{2}-\bar{\Delta}_{3}+\bar{\Delta}_{1}+\bar{\Delta}_{2}=
\nmb
&\sum_{i=1}^{r}\frac{1}{4(k_{i}+2)}((\bar{q}_{1i}-2w_{1i})^{2}+(\bar{q}_{2i}-2w_{2i})^{2}-(\bar{q}_{3i}-2w_{3i})^{2}-
\nmb
&(\bar{q}_{1i}-1)^{2}-\bar{q}_{2i}^{2}+(\bar{q}_{3i}-1)^{2}).
\nmb
\end{aligned}
\en
Using (\ref{A.conserv1}) we obtain
\be
\begin{aligned}
&\Delta_{3}-\Delta_{1}-\Delta_{2}-\bar{\Delta}_{3}+\bar{\Delta}_{1}+\bar{\Delta}_{2}=
\nmb
&\sum_{i=1}^{r}(\frac{w_{1i}(\bar{q}_{2i}-w_{2i})+w_{2i}(\bar{q}_{1i}-w_{1i})}{k_{i}+2}-\fr12\frac{\bar{q}_{2i}}{k_{i}+2}).
\nmb
\end{aligned}
\en
Hence, the mutual locality condition is
\begin{equation}
\sum_{i=1}^{r}\frac{w_{1i}(\bar{q}_{2i}-w_{2i})+w_{2i}(\bar{q}_{1i}-w_{1i})}{k_{i}+2}-\fr12(\bar{q}_{2}+\sum_{a}\bar{\lm}^{a}_{2}+\sum_{i=1}^{r}\frac{\bar{q}_{2i}}{k_{i}+2})\in\mathbb{Z}.
\end{equation}
The descendants which have been omitted in the vertex ${\cal{V}}^{(q_{2},\lm_{2})(\bar{q}_{2},\bar{\lm}_{2})}_{\bl_{2},\bar{\bt}_{2},\bw_{2}}(0)$ correct the expression above to
\begin{equation}
\sum_{i=1}^{r}\frac{w_{1i}(\bar{q}_{2i}-w_{2i})+w_{2i}(\bar{q}_{1i}-w_{1i})}{k_{i}+2}-\fr12(\bar{q}_{2}+\sum_{a}\bar{\lm}^{a}_{2}+\sum_{i=1}^{r}\frac{\bar{q}_{2i}}{k_{i}+2}+\bar{Q}_{2des})\in\mathbb{Z}.
\end{equation}
Therefore, the mutual locality condition is nothing else but the second line from (\ref{3.GSOB})):
\ber
\bar{q}_{2}+\sum_{a}\bar{\lm}^{a}_{2}+\sum_{i=1}^{r}\frac{\bar{q}_{2i}}{k_{i}+2}+\bar{Q}_{2des}\in2\mathbb{Z}.
\nmb
\enr
Similarly, one can prove the mutual locality of a vertex
operators of other types.


\end{document}